\documentclass[
 amsmath,amssymb,
 aps,
prx, twocolumn, longbibliography
]{revtex4}

\usepackage{graphicx}
\usepackage{dcolumn}
\usepackage{bm}
\usepackage{natbib}
\usepackage{color}
\graphicspath{{Pictures/}}

\newcommand{\Qvec}{\mathbf{Q}}
\newcommand{\nvec}{\mathbf{n}}

\begin{document}

\preprint{APS/123-QED}

\title{Uniaxial versus Biaxial Pathways in One-Dimensional Cholesteric Liquid Crystals}

 \author{Yucen Han}
\affiliation{Department of Mathematics and Statistics, University of Strathclyde, Glasgow, United Kingdom}
 
\author{James Dalby}%
\affiliation{Department of Mathematics and Statistics, University of Strathclyde, Glasgow, United Kingdom}%

\author{Benjamin MGD Carter}
\affiliation{H.H.~Wills Physics Laboratory and Bristol Centre for Functional Nanomaterials, University of Bristol, Tyndall Avenue, Bristol BS8 1TL, United Kingdom}
 
 \author{Apala Majumdar} \email[Email: ]{apala.majumdar@strath.ac.uk}%
\affiliation{Department of Mathematics and Statistics, University of Strathclyde, Glasgow, United Kingdom}%

 \author{Thomas Machon} \email[Email: ]{t.machon@bristol.ac.uk}%
\affiliation{H.H.~Wills Physics Laboratory, University of Bristol, Tyndall Avenue, Bristol BS8 1TL, United Kingdom}%



\date{\today}

\begin{abstract}

Cholesteric liquid crystals exhibit great morphological richness of static metastable states. Understanding the transitions between such states is key for the development of switchable devices. We show, using a quasi-one-dimensional model, that cholesterics exhibit distinct uniaxial and biaxial pathways between distinct minima. We study transitions between different layer numbers and prove, and show, that transition states are distinguished either through splay-mediated untwisting, understood through contact topology, or the presence of biaxiality. Furthermore we characterise a menagerie of additional saddle points that dictate the connectivity of the solution landscape.
\end{abstract}

\maketitle

Cholesteric liquid crystals display rich morphology, particularly in confinement. For a given geometry, with specified boundary conditions, one typically finds a huge variety of metastable states~\cite{tambovtsev2022topological}. These include arbitrarily knotted disclination lines~\cite{tkalec2011reconfigurable,tai2019three,machon2014knotted,machon2013knots}, torons~\cite{smalyukh2010three,chen2013generating} and hopf solitons~\cite{ackerman2017diversity}, inverse torons~\cite{eun2021layering}, point defect constellations~\cite{posnjak2017hidden,pollard2019point}, complex patterns in shells, cylinders and droplets~\cite{darmon2016topological,bouligand1984organization,darmon2016waltzing,tran2017change, emerson2018computing}, and Skyrmion lattices~\cite{fukuda2011quasi} amongst many others. This richness gives cholesteric liquid crystals great potential for the creation of soft devices. There is another key strength to cholesterics and all liquid crystal systems,  they flow (spontaneously in the case of active systems) and readily respond to applied fields~\cite{gartland2010electric}; in other words, they are switchable. For a switchable device, understanding and controlling pathways between stable states is just as important as understanding the stable states themselves, non energy-minimizing critical points play crucial roles in switching mechanics and selection of minimizers~\cite{hannonlinearity2020}. In a cholesteric system, chirality leads to many complex metastable configurations and one expects similar richness in the space of transition pathways between metastable states.

Many liquid crystalline configurations are distinguished by their topology, using homotopy theoretic invariants~\cite{mermin1979topological,alexander2012colloquium,poenaru1977crossing}. However, hese invariants often cannot distinguish between two cholesteric configurations that appear qualitatively distinct. We consider the simplest example in this letter~\cite{machon2017contact} -  the number of layers in a cholesteric cell, with all even (or odd) layer numbers equivalent from a homotopy-theoretic perspective. An alternative approach uses ideas from contact topology~\cite{machon2017contact,pollard2019point,eun2021layering} to provide more sophisticated topological invariants. 
In this theory, the key role is played by the twist density; 
if the twist density is nowhere-zero, the liquid crystal configuration is labelled as being chiral. Using the tools of contact topology~\cite{geiges2008introduction}, one can rigorously define new invariants (e.g. the layer number~\cite{machon2017contact,eun2021layering,hu2021stability}) that can distinguish between chiral configurations, for example helices with different numbers of twists/layers. This gives a new heuristic understanding of the cholesteric solution landscape, which we explore in this letter. We investigate pathways between helices with different numbers of twists, and use sophisticated topological and numerical studies to discover a dichotomy -the helices or equivalently the different chiral sectors can be connected by``untwisted states'' or they can be connected by exotic biaxial defects, which avoid the untwisting. Which pathway is preferred or observed is a delicate question that depends on the geometry, temperature, and material properties, and careful investigations of the parameter space can allow us to steer these pathways and manipulate the corresponding solution landscapes. Our work is a first step in the theoretical demonstration of such dichotomies for the simplest model cholesteric system, and our tools will extend to more complex higher-dimensional cholesteric systems to navigate through the labyrinth of metastable states and the pathways between them.  


We study a cholesteric sample inside a cell of height $h$, 
$\Omega = \{(x, y, z) \in\mathbb{R}^3 | \; 0 \leq z \leq h\}$. We work in the powerful Landau-de Gennes (LdG) formalism, where the cholesteric state is described by the LdG $\bf Q$-tensor order parameter, with Cartesian components $Q_{ij}$. The LdG $\Qvec$-order parameter is a symmetric, traceless $3\times 3$ matrix \cite{deGennes} 
\[
\Qvec = \sum_{i=1}^{3} \lambda_i \mathbf{e}_i \otimes \mathbf{e}_i
\]
where the eigenvectors, $\mathbf{e}_i$, model the preferred material directions and the corresponding eigenvalues, $\lambda_i$, measure the degree of orientational order about $\mathbf{e}_i$. The eigenvector with the largest positive eigenvalue is the nematic director ${\bf n}$ \cite{deGennes}. A $\Qvec$-tensor is  uniaxial if it has two equal non-zero eigenvalues and biaxial if it has three distinct eigenvalues 
\cite{majumdar-2010-article}. In particular, $\Qvec$ is said to be quasi-uniaxial if it has non-zero eigenvalues and a well-defined director, and maximally biaxial if it has a single zero eigenvalue.

Assuming translational symmetry in the $x$ and $y$ directions, the equilibrium or physically observable cholesteric states correspond to minimizers of the non-dimensionalized cholesteric energy per unit area~\cite{fukuda2010cholesteric}
\begin{align}
\bar{F}(\Qvec)=\frac{ h F(\Qvec)}{K_0} & =\int_0^1 \biggl\{\frac{1}{4}(\overline{\nabla}\cdot\Qvec)^2+\label{eq:energy-nondimensional_Q}\\
& \frac{\eta}{4}\left|\overline{\nabla}\times\Qvec+2\sigma \Qvec\right|^2+\lambda \frac{f_b(\Qvec)}{C}\biggr\}~\mathrm{d}\bar{z}.\nonumber
\end{align}
The elastic energy density has splay/bend~\cite{notesplay} and twist contributions: $\eta = K_1/K_0$ is the ratio of the twist and splay{/bend} elastic constants and quantifies the elastic anisotropy;
 $\sigma/2\pi = h/p_0$ where $p_0$ is the pitch of the cholesteric, gives the number of $2\pi$ rotations of the director in the z-direction; and  $\lambda = \frac{h^2C}{K_0}$ is a measure of domain height/size with $C$ a material-dependent positive constant.  Using parameter values in \cite{priestly2012introduction} and   \cite{mottram2014introduction}, a typical value is $\eta=\frac{1}{2}$. The values of $\eta$ can be tailored by the material such as 5CB and PBG  \cite{dietrich2020extremely}.
 We omit bars from rescaled variables. The bulk energy density is
\begin{equation}
    f_b(\Qvec):=\frac{A}{2}\textrm{tr}\Qvec^2-\frac{B}{3}\textrm{tr}\Qvec^3+\frac{C}{4}(\textrm{tr}\Qvec^2)^2,\label{eq:bulk_LdG}
\end{equation}
and dictates the isotropic-nematic phase transition as a function of temperature. The variable $A=\alpha(T-T^*)$ is a rescaled temperature, with $T^*$ a characteristic temperature, and $B>0$ is material dependent constant ($C$ is as above)~\cite{deGennes}. 
Typical values for the liquid crystal MBBA are, $\alpha = -0.042\times 10^6 N/(K\cdot m^2)$, $B = 0.64\times 10^6 N/m^2$, and $C = 0.35\times 10^6 N/m^2$, as used in our numerical simulations. We work with low temperatures for which $A<0$, so the set of $f_b$ minimizers are uniaxial ${\bf Q}$-tensors, of the form
$\Qvec=s_+(\nvec\otimes \nvec-\mathbf{I}/3)$, for arbitrary unit-vectors fields or nematic directors, $\nvec$. Here 
$s_+=(B+\sqrt{{B^2}+24|A|C})/4C$. 

We impose Dirichlet boundary conditions on the top and bottom of the cell, given by ${\bf Q}(z=0) = {\bf Q}(z=h) = s_+ \left ({\bf e}_x \otimes {\bf e}_x - {\bf I}/3 \right )$, where ${\bf e}_x$ is a unit vector in the $x$-direction. This is equivalent to including a Rapini-Papoular surface energy in the limit of infinite surface anchoring. 
In the weak anchoring regime, the pathway may proceed via transitions at the boundary~\cite{tenishchev2019multiple, tenishchev2021hysteresis, barbero2018twist}.
We focus on helical director fields
$${\bf n}_w = (\cos( \pi w z/h), \sin ( \pi w z/h), 0),$$
with $w \in \mathbb{Z}$ the twisting or layer number, and define the associated uniaxial $\Qvec$-tensor to be $\Qvec_w = s_+ \left(\nvec_w\otimes \nvec_w - \frac{\mathbf{I}}{3} \right)$. Deep in the cholesteric phase, the cholesteric free energy will be minimized by a quasi-uniaxial $\Qvec$-tensor close to $\Qvec_w$ (with a small degree of biaxiality~\cite{wright1989crystalline}). 
Such a helical texture will satisfy the boundary conditions only if the pitch $p$ satisfies $p = h/|w|$. Otherwise the system encounters geometric frustration and one typically finds a number of metastable helices, with different twisting numbers $w$~\cite{kiselev2005twist}. 

\begin{figure}
\centering
        \includegraphics[width=\columnwidth]{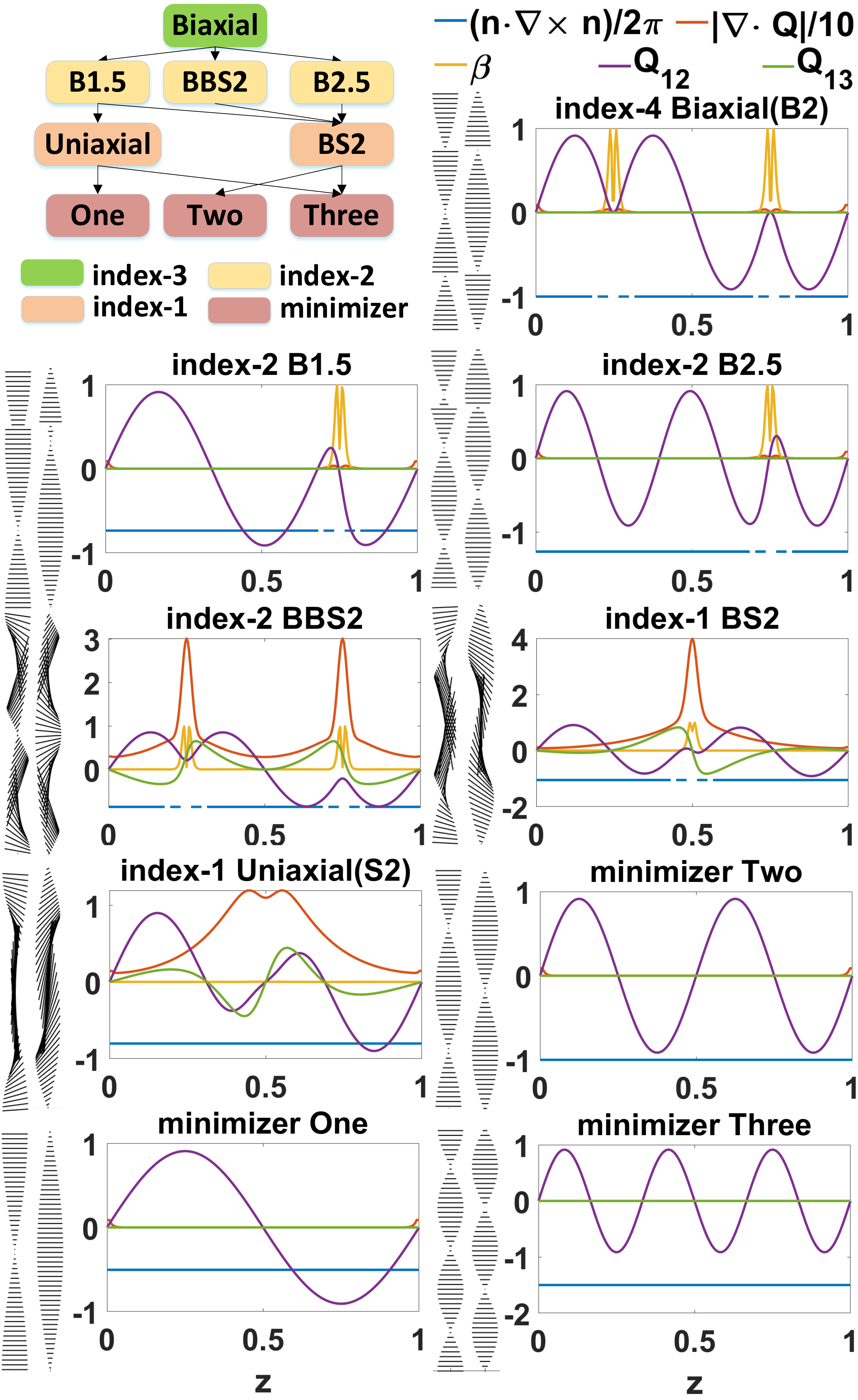}
        \caption{The solution landscape with $\lambda = 3500$, $\eta = 4.5$, $\sigma = 2\pi$, 
         $A = -B^2/3C$. 
         {We plot the twisting measure $(n\cdot\nabla\times n)/2\pi$ (represented by dash line near biaxial torus), the splay/bend measure $|\nabla\cdot\mathbf{Q}|/10$,
        the biaxiality measure $\beta = 1 - 6\frac{(tr \Qvec^3)^2}{|\Qvec|^6}$, the twisting in $xy$-plane $Q_{12}$ and the out of plane order $Q_{13}$.} We label the critical points as follows: `B' indicates the presence of biaxial tori, `S' indicates the presence of large splay, and the number at the end is the amount of rotation in the $xy$-plane, in units of $\pi$. The index-$4$ $Biaxial$($B2$) state has two biaxial tori and has unstable directions that tune the twisting periods and connect with the stable $One$ and $Three$ states. The $Biaxial$ state connects with three index-$2$ critical points, $B1.5$, $B2.5$, and $BBS2$. $B1.5$ and $B2.5$ have only one biaxial torus, with $3\pi/2$ and $5\pi/2$ twisting and connect $One$ and $Two$, and $Two$ and $Three$ respectively. $BBS2$ has a more complex profile with two biaxial tori, $2\pi$ twisting, and splay. The defect-free index-$1$ $Uniaxial$($S2$) state, is the transition state between the stable $One$ and $Three$ states. $BS2$ is the transition state between the stable $\Qvec_2$ and $\Qvec_3$.}
        \label{fig:solution_landscape}
\end{figure}

We are interested in pathways between different $\Qvec_w$, exploiting the full freedom of the $\Qvec$-tensor formalism. There is a qualitative difference between an $w-1 \to w$ transition and a $w-1 \to w+1$ transition. In the first case, the parity changes 
so that a pathway with fixed boundary conditions, will exhibit a defect in ${\bf n}$, or a point of maximal biaxiality in $\Qvec$. On the other hand, weak anchoring could allow for a defect-free pathway by deforming the texture at the boundary, as has recently been studied in the interesting papers \cite{ivanov2016energy,tenishchev2019multiple, tenishchev2021hysteresis}. In this case, the authors observe an interesting dichotomy between in-plane director slippage and an out-of-plane pathway, which has a visual resemblance to the uniaxial pathway observed here (Fig.~\ref{fig:solution_landscape}, $S2$), whether this connection has deeper significance remains to be seen.
For a $w-1 \to w+1$ transition, there is no analogous constraint and a defect-free quasi-uniaxial pathway is admissible, but there are hidden topological constraints of untwisting and splay along such quasi-uniaxial pathways, as demonstrated in the two theorems below. 

{{\bf Theorem 1:}} The director field ${\bf n}$ along any quasi-uniaxial pathway between ${\bf Q}_{w_1}$ and ${\bf Q}_{w_2}$, for $w_1 \neq w_2$, must untwist, that is ${\bf n} \cdot \nabla \times {\bf n}$ must vanish at some point in space and time along the pathway.

This is a refinement of a result given in~\cite{machon2017contact}, and is related to a number of results in contact topology~\cite{honda2000classification,yutaka1997classification,geiges2008introduction}. A proof is given in the SI. Topologically required untwisting can also be observed near point defects in cholesterics~\cite{ackerman2016reversal, pollard2019point}. 
Since cholesterics prefer to twist, with the term {$({\bf n} \cdot \nabla \times {\bf n}+\sigma)^2$} in the Frank free energy, the topologically required untwisting predicts a chiral energy barrier between different values of $w$. 

Next, consider a pathway with zero splay $(\nabla \cdot \Qvec = 0)$, such as a linear interpolation between ${\bf Q}_{\omega-1}$ and ${\bf Q}_{\omega+1}$. If we restrict ourselves to $\Qvec$-tensors with zero splay, then we necessarily have
\begin{equation}
Q_{13}^\prime = Q_{23}^\prime= Q_{33}^\prime=0.\label{eq:splay-constraints}
\end{equation}
Denoting the pathway by $\Qvec(t,z)$, we have $\Qvec(t = 0) = \Qvec_{\omega - 1}$ and $\Qvec (t = 1) = \Qvec_{\omega+1}$, with $Q_{13} = Q_{23}=0$ and $Q_{33} =-s_+/3<0$ for all $t\in [0,1]$, so that the nematic director is always in the $xy$-plane and is splay free. The leading order energy is
\begin{align}
    F(\Qvec)=F(\mathbf{q})=&\int_0^1\biggl\{\frac{\eta}{2}|\nabla \mathbf{q}|^2+ 2\sigma\eta(\mathbf{q}\cdot\nabla\times\mathbf{q}) +\nonumber\\
    &\frac{\lambda}{C}\tilde{f}_b(s_+,|\mathbf{q}|^2) + C_0(s_+,\eta,\sigma)\biggr\}~\mathrm{d}z,\label{eq:functional_q}
\end{align}
where $\mathbf{q} = \left(\frac{Q_{11}-Q_{22}}{2},Q_{12}\right)$ and $C_0$ is a function depending on $s_+$, $\eta$, and $\sigma$ i.e., there are only two degrees of freedom in the class of splay free configurations. 
\eqref{eq:functional_q} bears considerable similarity to a one-dimensional model studied by Golovaty et al.~\cite{golovaty2021one}. In the reduced description \eqref{eq:functional_q}, the boundary conditions are ${\bf q} = (s_+/2,0)$ at the top and bottom of the cell. In the helical state ${\bf Q}_w$, the vector ${\bf q}$ winds by $2w \pi$ between the top and bottom.  The change in winding number across the cell height forces a zero in ${\bf q}$, and points of maximal biaxiality correspond to $|{\bf q}| = s_+/6$. Thus, any splay-free pathway must cross the circle of maximal biaxiality as stated below:

{\bf Theorem 2} Any smooth zero-splay {and bend} pathway ($\nabla \cdot {\bf Q}=0$) between ${\bf Q}_{w_1}$ and ${\bf Q}_{w_2}$, $w_1 \neq w_2$, must contain points of maximal biaxiality.

A full proof, and slight generalisation, is given in the SI. Theorem 1 implies that the director field along any quasi-uniaxial pathway must untwist. Theorem 2 implies any quasi-uniaxial pathway must also have non-zero splay/bend. This reflects the large number of complex compatibility conditions~\cite{pollard2021intrinsic,da2021moving,selinger2021director} that relate the different deformation modes (splay, twist, bend, biaxial splay~\cite{selinger2021director,selinger2018interpretation,machon2016umbilic}) of a director field. We note this as the twist elastic constant can be substantially smaller than the splay or bend constants~\cite{dietrich2020extremely}. Combined, these two Theorems show that any pathway that changes the twisting number can either introduce defects (maximal biaxiality) or untwist (or both), and this conclusion applies to any pathway, not just pathways between minimizers of the free energy but also to pathways between distinct non-equilibrium states in hydrodynamic flows, where energy principles may not exist~\cite{tang2020minimization,pieranski2020tropisms,vcopar2020microfluidic}.

In Figure~\ref{fig:solution_landscape}, we illustrate this dichotomy in pathways between $\Qvec_\omega$, by looking at the critical points of the energy \eqref{eq:energy-nondimensional_Q} for a large cell with $\lambda =3500$ and $\eta = 4.5$. In the $\lambda \to \infty$ limit, the bulk energy dominates and energy minimizers tend to be approximately uniaxial i.e. solutions of the Euler-Lagrange equations in the restricted space of bulk energy minimizers \cite{majumdarzarnescu2010}. The leading order minimizer profiles are the helical states $\Qvec_\omega$, with $\omega=\frac{\sigma}{\pi}$.
We fix $\sigma = 2\pi$ in what follows and focus on (transition) pathways between three locally stable critical points of the free energy in Figure~\ref{fig:solution_landscape}: $\mathbf{Q}_{1}$ ($One$, $\pi$ twisting), $\mathbf{Q}_{2}$ ($Two$, $2\pi$ twisting) and $\mathbf{Q}_3$ ($Three$, $3\pi$ twisting). The $One$ and $Three$-helical states have the same energy for any $\lambda$, $\eta$, and $K_0$, whilst $\Qvec_{2}$ has the minimum energy amongst the three helical states.

To compute these pathways, we numerically compute solutions of the Euler-Lagrange equations for \eqref{eq:energy-nondimensional_Q}, which are critical points of the free energy. The Morse index of a critical point is the number of negative eigenvalues of the corresponding Hessian of the free energy, evaluated at the critical point \cite{milnor1963morse}. Minimizers (local and global) have index-$0$, index-$k$ saddle points are unstable in $k$ distinguished eigendirections. We use the high-index optimization-based shrinking dimer (HiOSD) method \cite{yin2019high} and upward/downward search algorithms \cite{yin2020construction}, to compute sample solution landscapes in Figure~\ref{fig:solution_landscape} (see SI).

To find novel biaxial critical points, we construct an initial condition with $2\pi$ twisting in the $xy$-plane, but with two discontinuous points of the nematic director, where {$\mathbf{q} = 0$} (see SI). Using Newton's method and this initial condition, we can obtain a splay-free critical point, labelled as the $Biaxial$ solution {(see Fig. \ref{fig:solution_landscape})}, for various $\lambda$ and $\eta$. 
The $Biaxial$ solution has two biaxial tori around the discontinuous points, and each biaxial torus contains two points of maximal biaxiality and a point of uniaxiality at the center. The biaxial torus is a universal defect structure arising from topological constraints \cite{henao2017uniaxial}, and can be understood in part {by the proof of Theorem 2 in SI}. 
For sufficiently small $K_0$ (when splay is not heavily penalised), the $Biaxial$ solution is an unstable saddle point (index $>1$), and the Morse index of the $Biaxial$ state increases as $\lambda$ increases.

With the $Biaxial$ state as the parent state (state with highest index), we compute the solution landscape for {$\lambda = 3500$ (corresponding to a small cell of about 0.2$\mu m$ height)}, and $\eta =4.5$, in Fig. \ref{fig:solution_landscape}. The computation reveals a menagerie of critical points, which we classify broadly as either {quasi-uniaxial or biaxial (without and with maximal biaxiality)
and with and without splay (non-zero and zero $|\nabla\cdot\mathbf{Q}|$ respectively)}. We observe, in passing, that the index of the biaxial critical points is twice the number of biaxial tori in the splay-free case, and equal to the number of biaxial tori with splay in Figure~\ref{fig:solution_landscape}. Notably, we find two competing pathways between the stable uniaxial $One$ and $Three$-helical states. According to Figure~\ref{fig:solution_landscape}, there is the uniaxial pathway via the index-$1$ $Uniaxial$ state, and the biaxial pathway via the index-$4$ $Biaxial$ state. Theorem 1 implies that the uniaxial pathway must untwist (i.e. have ${\bf n} \cdot \nabla \times {\bf n}=0$ at some point and, as suggested by Theorem 2, this untwisting is mediated by splay. On the other hand, the $Biaxial$ state, $B2$, is splay-free and indeed the entire pathway in this case can be chosen in the class of splay-free configurations \eqref{eq:functional_q}. 

Recent work has found similar structures to {(half of)} our uniaxial pathway in boundary-driven unwinding transitions, this suggests that the uniaxial critical point may become further preferred under an applied electric field, beyond the Fr\'{e}edericksz transition~\cite{tenishchev2019multiple, tenishchev2021hysteresis}. Such out-of-plane pathways may be also be relevant in the study of helical axis transitions~\cite{zheng2016three}. A related set of uniaxial untwisting pathways in nematics were considered by Goldbart and Ao~\cite{goldbart1990intrinsic, goldbart1991intrinsic}.

The question then is -- which pathway is preferred? This then depends on $\eta$, $\lambda$ and the temperature.  
In the $K_0\to\infty$ limit,  \eqref{eq:energy-nondimensional_Q}, only the splay-free states: $Biaxial(B2)$, $B1.5$, $B2.5$, $One$, $Two$, and $Three$,   
survive as critical points. In particular, the $Biaxial$ state is always a critical point of \eqref{eq:energy-nondimensional_Q} for all $K_0$, and the index of $Biaxial$ state decreases from $4$, to $3$, to $2$, to $0$, as $K_0$ increases from $1$ to $\infty$. The $Uniaxial$($S2$), $BS2$, and $BBS2$, have splay; their Morse indices and energy increase with $K_0$. 
{As temperature decreases, uniaxiality is energetically preferred to biaxiality. The $Biaxial(B2)$, $B1.5$, $B2.5$, $BS2$, and $BBS2$ critical points have points of maximal biaxiality, they have increasing energy and Morse indices with decreasing temperature, in contrast to the $Uniaxial$($S2$), $One$, $Two$, and $Three$ which have decreasing energy with decreasing temperature.}

\begin{figure}
\centering
        \includegraphics[width=\columnwidth]{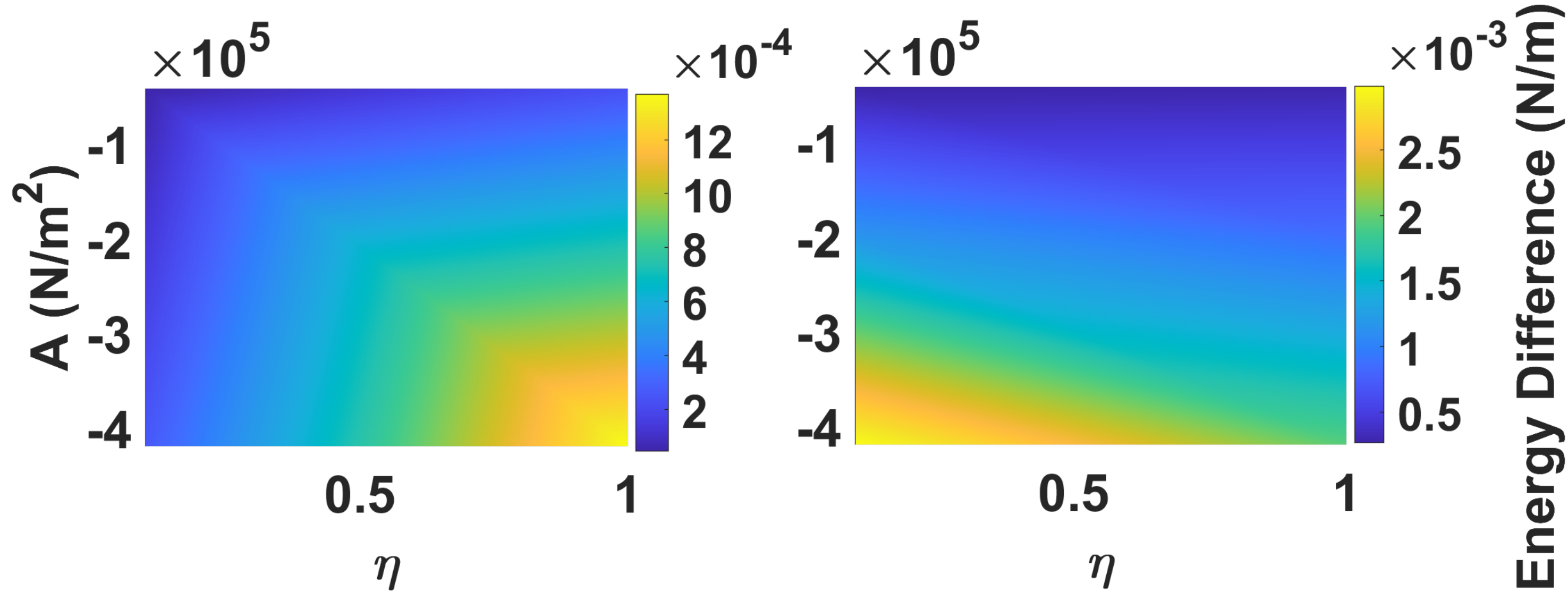}
        \caption{{The difference between energy barriers (the energy $F$ before non-dimensionalization in \eqref{eq:energy-nondimensional_Q}) of the uniaxial and biaxial pathways as a function of the re-scaled temperature, $A$ and elastic anisotropy $\eta$. Left: small cell with $\lambda = 3500$ ($h \approx 0.2\mu m$); Right: large cell with $\lambda = 17500$ ($h \approx \textcolor{red}{1.4}\mu m$).}}
        \label{fig:energy_barrier}
\end{figure}

Returning now to the choice between the biaxial ($B2$) and uniaxial ($S2$) pathways, the energy barrier of the pathway is the energy difference between either the $Uniaxial (S2)$ or $Biaxial (B2)$ critical point, and the stable  $One$ (or $Three$) states. {The difference between two energy barriers is equal to the energy difference between $Uniaxial (S2)$ and $Biaxial (B2)$.
When the temperature is high, the energy difference is low since uniaxiality is only weakly preferred. 
The uniaxial pathway cannot be observed for extreme parameter values, such as large $K_0$. More importantly, for the range of $A$ and $\eta$ considered in Figure~\ref{fig:energy_barrier}, for small and large $\lambda$, the uniaxial pathway is always energetically preferable but the uniaxial and biaxial pathways have comparable energy barriers, so that experimental observations of both pathways are possible in practice.}


What are the implications for experiments? During coarsening or other dynamic processes (for example, the Helfrich-Hurault instability~\cite{blanc2021helfrich}) cholesterics will typically undergo morphological changes whereby `layer numbers' are observed to change in a discrete fashion. Our work suggests that a cholesteric system can choose to mediate such a transition by either untwisting via splay and bend distortions, or through biaxiality (defects). Experimentally verifiable optical signatures for these states as well as their pathways are shown in the SI. Our findings illustrate that biaxiality may become experimentally observable for these toy systems, with {high temperature} and large $K_0$, either as equilibria or in pathways between distinct equilibria. Further, we conjecture that we can construct hierarchies of LdG critical points from building points e.g. the biaxial torus or untwisting states (the work of Goldbart and AO suggests that the barriers we study increase with the number of twists~\cite{goldbart1990intrinsic, goldbart1991intrinsic}). 
In higher dimensions, cholesterics exhibit great morphological richness, knotted disclination lines~\cite{tkalec2011reconfigurable,tai2019three,machon2014knotted,machon2013knots} and hopf solitons~\cite{ackerman2017diversity}, for example. Whether the ideas discussed in this letter can be used to study transition pathways for these complex problems, remains an interesting open question (we note biaxial tori are expected to be minimizing for certain problems~\cite{henao2017uniaxial}), and our major challenge is to discover universal principles for solution landscapes of confined soft matter systems.

Taxonomy: TM and BMGDC proposed the problem, TM conceived and proved Theorems 1 and 2. AM led the asymptotic analysis and second part of the manuscript. YH performed the numerical simulations; YH and JD carried out the analysis; BMGDC performed initial numerical simulations. YH, JD, BMGDC, AM and TM contributed to the manuscript writing and revision.

Acknowledgements: AM, JD and YH gratefully acknowledge support from a Heilbronn Small Grant that facilitated this collaboration. 
YH is supported by a Royal Society Newton International Fellowship. The authors thank Yiwei Wang for sharing computational code for optical profiles. TM and BMGDC acknowledge funding from EPSRC with grant code EP/L016648/1.

\section{Supplementary Information}

\section{Proof of Theorem 1}

This theorem uses ideas of contact topology, which is the study of the topology of contact structures, see Geiges' book~\cite{geiges2008introduction} for a thorough introduction. In a three-dimensional space $\Omega$, a contact structure is defined by a one-form $\alpha$, in coordinates $\alpha_i$, which satisfies 
$$ \alpha \wedge d \alpha \neq 0,$$
the 3-form $\alpha \wedge d \alpha$ is equal to $\epsilon^{ijk}\alpha_i \partial_j \alpha_k dV$. The contact structure itself is defined as the collection of 2-planes containing tangent vectors which are orthogonal to $\alpha$. If $\alpha(x)_i$ is the value of $\alpha$ at $x \in \Omega$, the orthogonal 2-plane at $x$ is defined as all vectors $v^i$ satisfying $v^i \alpha_i(x)=0$. If we convert $\alpha_i$ to a vector field $A^i$ using a Riemannian metric to raise indices, then the condition  $\alpha \wedge d \alpha \neq 0$ becomes 
$${\bf A} \cdot \nabla \times {\bf A} \neq 0,$$
which is just the condition for non-zero twist. We now state and prove theorem 1. This a refined version of the result that appears in Ref.~\cite{machon2017contact}. Those looking for a more intuitive picture are referred to the arguments of Ref.~\cite{machon2017contact}.

{{\bf Theorem 1:}} The director field ${\bf n}$ along any quasi-uniaxial pathway between ${\bf Q}_{w_1}$ and ${\bf Q}_{w_2}$, for $w_1 \neq w_2$, must untwist, that is ${\bf n} \cdot \nabla \times {\bf n}$ must vanish at some point in space and time along the pathway.

Recall we define ${\bf n}_w = (\cos( \pi w z/h), \sin ( \pi w z/h), 0),$ with $w \in \mathbb{Z}$. For the purposes of this theorem we assume the domain $\Omega$ is periodic in $x$ and $y$ (with arbitrarily large period). If we take open boundary conditions, so that $\Omega \cong \mathbb{R}^2 \times [0,1]$, then the result can still be established using ideas of proper isotopy (see work of Eliashberg~\cite{eliashberg1993classification} as well as the work of J.J. Tripp~\cite{tripp2006contact}), which requires some physically-reasonable caveats on the allowed deformations.

Proof: suppose there is a pathway that does not satisfy the above property, then we have a director field satisfying ${\bf n} \cdot \nabla \times {\bf n} \neq 0$ for all $(x, t) \in \Omega \times [0,1]$. Using the Euclidean metric we can identify ${\bf n}$ with a 1-form $\alpha$, which then satisfies $\alpha \wedge d \alpha \neq 0$. Hence ${\rm ker}(\alpha)$ defines a contact structure on $\Omega$ for each $t \in [0,1]$. Write $\alpha_w$ for the 1-form dual to ${\bf n}_w$. Our assertion is then equivalent to the existence of an isotopy of contact structures $\alpha(t)$, $0 \leq t \leq 1$, with $\alpha(0) = \alpha_{w_1}$ and $\alpha(1) = \alpha_{w_2}$, satisfying the boundary conditions $\alpha = dx$ on the top and bottom surfaces, and where we define
$$\alpha_w = \cos( \pi w z/h) dx + \sin ( \pi w z/h)dy$$
Suppose such a pathway $\alpha(t)$ exists. Our boundary conditions on the top and bottom of $\Omega$ are identical. We can therefore glue the top and bottom together together to obtain a family of 1-forms $\tilde \alpha(t)$ on the 3-torus $T^3$. This family will be of class $C^\infty$ everywhere except the gluing site, where it is $C^0$ in the direction across the gluing surface (although it is $C^\infty$ at $t=0$ and $t=1$). By a small perturbation, we may make $\tilde \alpha(t)$ smooth. Hence any pathway $\alpha(t)$ defines a smooth pathway $\tilde \alpha(t)$ between $\alpha_{w_1}$ and $\alpha_{w_2}$ on $T^3$. But by the theorem of Yutaka~\cite{yutaka1997classification}, such a pathway cannot exist. Hence the pathway $\alpha(t)$ also cannot exist.

\section{Proof of Theorem 2}

{\bf Theorem 2} Any smooth zero-splay $(\nabla \cdot {\bf Q}=0)$ pathway between ${\bf Q}_{w_1}$ and ${\bf Q}_{w_2}$, $w_1 \neq w_2$, must contain points of maximal biaxiality.

Proof:
In our notation, any splay-free $\Qvec$-tensor matching the boundary conditions may be described by a vector ${\bf q}(z)=(q_1(z), q_2(z))$ as
$$\Qvec = \begin{pmatrix} \frac{s_+}{6} + q_1 & q_2 & 0 \\
q_2 & \frac{s_+}{6} - q_1 & 0 \\
0 & 0 & -\frac{s_+}{3} \end{pmatrix}.
$$
Satisfying the boundary conditions requires ${\bf q}(0) = {\bf q}(h) = {(s_+/2,0)}$. The helical state ${\bf Q}_w$ corresponds to $${\bf q}_w(z) = \frac{s_+}{2} (\cos ( 2\pi w z/h), \sin ( 2\pi w z/h)),$$ this defines a closed loop in the $(q_1, q_2)$ plane of radius $s_+ / 2$, winding around the origin $w$ times. Any one-parameter family of splay-free $\Qvec$-tensors matching the boundary conditions can be described by a 1-parameter family of loops ${\bf q}(z, t)$, $0 \leq t \leq 1$, such that ${\bf q}(z,0) = {\bf q}_{w_1}$, ${\bf q}(z,1) = {\bf q}_{w_2}$, ${\bf q}(0,t) = {\bf q}(h,t) = (s_+/2,0)$ for all $t$. Now since different ${\bf q}_w$s have different winding numbers around the origin in the $(q_1, q_2)$-plane, any pathway between ${\bf q}_{w_1}$ and ${\bf q}_{w_2}$ must pass through the origin, so there is at least one $z^\ast$ and $t^\ast$ such that ${\bf q}(z^\ast, t^\ast) = 0$. 

Now, a $\Qvec$-tensor is maximally biaxial if its determinant vanishes, and a short calculation shows that this happens in our case when $\sqrt{q_1^2+q_2^2} = s_+/6$, defining a circle of radius $s_+/6$ in the $(q_1, q_2)$ plane. Since the initial and final states ${\bf q}_{w_1}$ and ${\bf q}_{w_2}$ lie outside this circle, and any pathway must pass through the origin which is inside the circle, the pathway must pass through the circle of maximal biaxiality. Hence the pathway must contain points of maximal biaxiality.

As an extension, we note that ${\bf Q}_{w_1}$ and ${\bf Q}_{w_2}$ may be replaced by quasi-uniaxial splay-free $\Qvec$-tensors which correspond to closed loops in the $(q_1, q_2)$ plane outside the circle of maximal biaxiality, winding around the origin $w_1$ and $w_2$ times (respectively).

\begin{figure*}[h!t]
\centering
        \includegraphics[width=1.5\columnwidth]{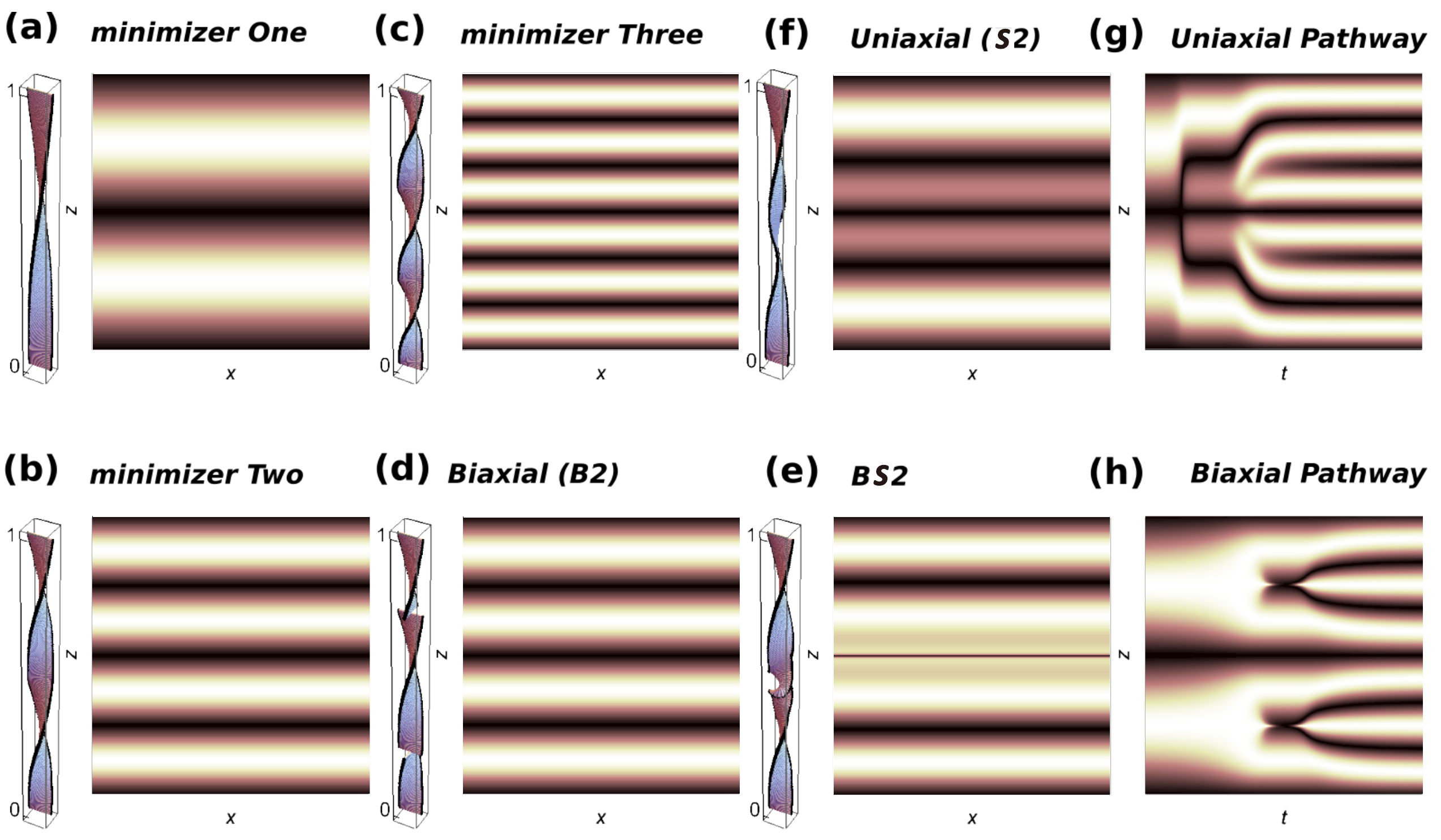}
        \caption{Optical signatures for a set of metastable states (a), (b), (c); and critical states (f), (d), (e). (g) (h) show space-time optical signatures for the uniaxial and biaxial pathways.}
        \label{fig:optical_modelling}
\end{figure*}

\section{Computational Methods}

We construct the initial condition for the $Biaxial(B2)$ state as follows
\begin{align}
    &\Qvec(z)=s_+\left(\mathbf{n}\otimes\mathbf{n}-\frac{1}{3}\mathbf{I}\right)\nonumber\\
    &=\begin{cases}
    \mathbf{Q}_{2}(z),\ z\in(0,0.25)\cup(0.75,1)\\
    S\Qvec_2(z)S^T,\ z\in(0.25,0.75),
    \end{cases}\nonumber
\end{align}
where $S$ is the rotation matrix which rotates the state on $z\in(0.25,0.75)$ by $90$ degree in the $xy$-plane.
The initial condition has $2\pi$ twisting, but with two discontinuous points of the nematic director $\mathbf{n}$, at $z = 0.25$ and $0.75$, and the director flips by $90$ degrees at these points.

The high-index optimization-based shrinking dimer (HiOSD) method \cite{yin2019high} is 
a local-search algorithm for the computation of saddle point of arbitrary indices, which can be viewed as a generalization of the optimization-based shrinking dimer method for searching index-$1$ saddle points \cite{zhang2016optimization}. It is an efficient tool for constructing the solution landscape searching from high- or low-index saddle points and revealing the connectivity of saddle points and minimizers.

For a non-degenerate index-$k$ saddle point $\hat{\mathbf{Q}}$, the Hessian $\mathbb{H}(\mathbf{Q})=\nabla^2 F(\mathbf{Q})$ at $\hat{\mathbf{Q}}$ has exactly $k$ negative eigenvalues $\hat{\lambda}_1\leqslant \cdots \leqslant\hat{\lambda}_k$ with corresponding unit eigenvectors $\hat{\mathbf{v}}_1,\ldots, \hat{\mathbf{v}}_k$ satisfying $\big\langle\hat{\mathbf{v}}_j, \hat{\mathbf{v}}_i\big\rangle = \delta_{ij}$, $1\leqslant i, j \leqslant k$.
Define a $k$-dimensional subspace $\hat{\mathcal{V}}=\mathrm{span}\big\{\hat{\mathbf{v}}_1,\ldots, \hat{\mathbf{v}}_k\big\}$, then $\hat {\mathbf{Q}}$ is a local maximum on a $k$-dimensional linear manifold $\hat{\mathbf{Q}}+\hat{\mathcal{V}}$ and a local minimum on $\hat{\mathbf{Q}}+\hat{\mathcal{V}}^\perp$, where $\hat{\mathcal{V}}^\perp$ is the orthogonal complement space of $\hat{\mathcal{V}}$. 

The $k$-dimensional subspace $\hat{\mathcal{V}}$ is constructed by solving the following constrained optimization problem using gradient flow,
\begin{equation}\label{minvi}
\min_{\mathbf{v}_i\in\mathbb{R}^n} \quad\langle \mathbb{H}(\mathbf{Q})\mathbf{v}_i, \mathbf{v}_i\rangle,\quad
\mathrm{s.t.} \quad\langle \mathbf{v}_j,\mathbf{v}_i\rangle=\delta_{ij},\quad j=1,2,\ldots,i.
\end{equation}
The $k$-index $\hat{\mathbf{Q}}$ can be achieved by a minimax optimization problem
\begin{equation}
\min_{\mathbf{Q}_{\hat{\mathcal{V}}^{\perp}}\in\hat{\mathcal{V}}^{\perp}}\max_{\mathbf{Q}_{\hat{\mathcal{V}}}\in\hat{\mathcal{V}}} F(\mathbf{Q}_{\hat{\mathcal{V}}^{\perp}} + \mathbf{Q}_{\hat{\mathcal{V}}}),
\end{equation}
where $\mathbf{Q}_{\hat{\mathcal{V}}} = \mathcal{P}_{\hat{\mathcal{V}}}\mathbf{Q}$ is the orthogonal projection of $\mathbf{Q}$ on $\hat{\mathcal{V}}$, and $\mathbf{Q}_{\hat{\mathcal{V}}^{\perp}} = \mathbf{Q} - \mathcal{P}_{\hat{\mathcal{V}}}\mathbf{Q}$.

Thus with the simultaneous Rayleigh-quotient iterative minimization method, a dynamical system of the HiOSD is developed for finding an index-$k$ saddle point, 
\begin{equation}\label{dynamics}
\left\{
\begin{aligned}
\beta^{-1}\dot{\mathbf{Q}}   & =- \left(\mathbf{I}-2\sum_{j=1}^{k}\mathbf{v}_j \mathbf{v}_j^\top\right)\nabla F(\mathbf{Q}), \\
\gamma^{-1}\dot{\mathbf{v}}_i & = -\left(\mathbf{I}-\mathbf{v}_i\mathbf{v}_i^\top-2\sum_{j=1}^{i-1}\mathbf{v}_j \mathbf{v}_j^\top\right)\mathbb{H}(\mathbf{Q})\mathbf{v}_i,\; i=1,\ldots,k,
\end{aligned}
\right.
 \end{equation}
where the state variable $\mathbf{Q}$ and $k$ direction variables $\mathbf{v}_i$ are coupled, $\mathbf{I}$ is the identity operator and $\beta,\gamma>0$ are relaxation parameters. 

{To construct a pathway, we first calculate all the unstable directions which are eigenvectors of the critical/saddle point (Uniaxial or Biaxial state), corresponding to the negative eigenvalues of the Hessian of the free energy.
We perturb the critical/saddle point a bit along the positive or negative unstable direction. We get a new pathway between the One and Three twist helical states if the critical/saddle point converges to the One and Three twist helical-states directly, with opposite perturbations of the unstable eigendirections in the gradient flow dynamics, without crossing another critical point.}

\section{Optical Modelling}

In order to predict the optical motifs, we use the Jones Matrix formalism; which considers the theoretical optical transmission of a plane wave of normal incidence through a twisted nematic material.  The incident wave is modelled as propagating in the $\bf{e}_y$ direction.  The cholesteric is considered as a collection of nodes with a specific magnitude of birefringence depending on the local orientation of the director field with respect to the incident wave. Using the $2\times 2$ Jones Matrix formalism, we calculate the output of the field, $E^o$ as
\begin{equation}
\begin{pmatrix}
E_x^o \\
E_z^o
\end{pmatrix} = \begin{pmatrix}
0 & 0\\
0 & 1
\end{pmatrix}\bf{J}
\begin{pmatrix}
1 & 0\\
0 & 0
\end{pmatrix}\begin{pmatrix}
E_x^i \\
E_z^i
\end{pmatrix}
\end{equation}
where $E^i$ is the magnitude of the incident wave in the specified directions and $\bf{J}$ is the Jones matrix as given by
\begin{equation}
\bf{J} = \begin{pmatrix}
\cos ^2(\gamma)\mathrm{e}^{i\phi_e} + \sin ^2(\gamma)\mathrm{e}^{i\phi_0} & \cos(\gamma)\sin(\gamma)(\mathrm{e}^{i\phi_e} - \mathrm{e}^{i\phi_0})\\
\cos(\gamma)\sin(\gamma)(\mathrm{e}^{i\phi_e} - \mathrm{e}^{i\phi_0}) & \sin ^2(\gamma)\mathrm{e}^{i\phi_e} + \cos ^2(\gamma)\mathrm{e}^{i\phi_0}
\end{pmatrix}.
\end{equation}
The wavelength of the incident beam, $\lambda$, is considered to be $\lambda = 450$nm, $\gamma$ represents the azimuthal angle between the extraordinary axis ($\bf{e}_x$) and the alignment of the director, and $\phi_0$ and $\phi_e$ are the ordinary and extraordinary phase shifts, respectively, induced by the nematic where
\begin{align}
\phi_0 &= \frac{2\pi}{\lambda}n_0
\\\phi_e &= \frac{2\pi}{\lambda}n_e(\beta).
\end{align}
In this case we consider $n_0=1.52$ being the ordinary refractive index of the cholesteric and $n_e(\beta)$ being the local extraordinary index determined by
\begin{equation}
n_e ^2 (\beta) = \frac{n_0 n_e}{\sqrt{n_0^2 \sin^2(\beta) + n_e^2 \cos ^2(\beta)}}
\end{equation}
which is dependent upon the extraordinary refractive index of the material; in this case $n_e = 1.90$; and $\beta$ being the angle between the incident wave ($\bf{e}_y$ direction) and the local director.

In experiments, these transmission profiles are studied via optical microscopy \cite{doi:10.1142/2337}. The numerically computed transmission profiles for the uniaxial critical points, with splay, are distinguished via distinct motifs with broadly distributed peaks that are fractured by localised `dark' regions. We note that the optical signature of the \textit{Biaxial} state ($B2$) is indistinct from the optical signature of the doubly twisted metastable state (Minimizer Two), possibly because the Jones matrix is inherently uniaxial and cannot capture biaxiality.

\bibliographystyle{unsrt}
\bibliography{twisting_arxiv_0815}

\end{document}